# NOVEL TEXT CATEGORIZATION BY AMALGAMATION OF AUGMENTED k-NEAREST NEIGHBOURHOOD CLASSIFICATION AND k-MEDOIDS CLUSTERING

RamachandraRao Kurada[1], Dr. K Karteeka Pavan[2], M Rajeswari[3] and M Lakshmi Kamala[3]

[1]Research Scholar (Part-time), AcharyaNagarjuna University, Guntur
Assistant Professor, Department of Computer Applications
Shri Vishnu Engineering College for Women, Bhimavaram
[2]Professor, Department of Information Technology
RVR & JC College of Engineering, Guntur
[3]Department of Computer Applications
Shri Vishnu Engineering College for Women, Bhimavaram

*Abstract*

*Machine learning for text classification is the underpinning of document cataloging, news filtering, document steering and exemplification. In text mining realm, effective feature selection is significant to make the learning task more accurate and competent. One of the traditional lazy text classifier k-Nearest Neighborhood (kNN) has a major pitfall in calculating the similarity between all the objects in training and testing sets, there by leads to exaggeration of both computational complexity of the algorithm and massive consumption of main memory. To diminish these shortcomings in viewpoint of a data-mining practitioner an amalgamative technique is proposed in this paper using a novel restructured version of kNN called AugmentedkNN(AkNN) and k-Medoids(kMdd) clustering.The proposed work comprises preprocesses on the initial training set by imposing attribute feature selection for reduction of high dimensionality, also it detects and excludes the high-fliers samples in the initial training set and restructures a constrictedtraining set. The kMdd clustering algorithm generates the cluster centers (as interior objects) for each category and restructures the constricted training set with centroids. This technique is amalgamated with AkNNclassifier that was prearranged with text mining similarity measures. Eventually, significantweights and ranks were assigned to each object in the new training set based upon their accessory towards the object in testing set. Experiments conducted on Reuters-21578 a UCI benchmark text mining data set, and comparisons with traditional kNNclassifier designates the referredmethod yieldspreeminentrecitalin both clustering and classification.*

## 1. Introduction

Pattern recognition concord with bestowing categories to samples, are declared by a set of procedures alleged as features or characteristics. Despite of prolific investigations in the past decannary, and contemporary hypothesis with extemporized thoughts, suspicion and speculation that exists, pattern recognition is facing confront in resolving real world problems [1].There are two major types of pattern recognition problemscalled unsupervised and supervised. In classification or the supervised learning, each sample in the dataset appears with a pre-designated category label, where as in clustering or the unsupervised learning, there is no pre-designated category label and assemblage is done based on similarities. In this paper, a combinational learningapproach using classification and clustering is projected for the recognition of text classifier accuracy and adequacy in judgment.

With the growth of the internet, efficient and accurate Information Retrieval (IR) systems [2]are of great importance. The present day search engines are able to accomplish enormous amount of data concerned in Exabyte's. Text classification effort involuntarily to decide whether adocument or part of a document has exacting distinctivenesstypically predestined on whether the documentenclose aconvinced type of topic or not. Uniquely, the topic of fascination is not defined absolutely or more accurately by the users and as a substitute they are arranged with a set of documents that comprise the enthrallment of both (positive and negative training set). The decision-making procedure [3] despotically hauls out the features of text documents and helps to discriminate positive from negatives. It also administers those appearances inevitably to sample documents with the help of text classification systems. The k-nearest neighbor rule is a simple and effective classifier for document classification. In this technique, a document is fashionable into an exacting category if the category has utmost diversity pertaining to the k nearest neighbors of the documents in the training set. The k nearest neighbors of a test document isordered based on their content similarity with the documents in the training set.

This paper probes and recognizes the payback of amalgamation of clustering and classification for text categorization. This paper is prearranged as follows: Section 2 gives an abrupt analysis of antecedent works on text classification. Section 3 is devoted to present the proposed approach for document classification. In Section 4 extensive experimental evaluation on text corpus derived from the Reuters-21578, a real time data set is summarized and analyzed. The resulting text categorization rates are comparing favorably for the referred methods with those of the standard method. Finally, conclusion is drawn and future research directions are identified in Section 5.

## 2. Literature Survey

TM Cover and PE Hart first proposed an elementary and straightforward classification method called the K-Nearest Neighbor (kNN) [4], where no former awareness about the circulation of the data is needed and neighborhood estimation was based on k value.kNN is a non-parametric lazy learning algorithm, since it does not make any postulations on the original data allocation and there is no precise training phase to do any simplification over the data points. The fascination of kNNdominion was of disunion in its minimal training phase and a lavish testing phase.

Few topological methods that pilot to the current state-of-the-are wKNNmethod intended by TBaileyand AK Jain [5] is aimproved version of simple kNN, assigns weights to training points according to the distance from centroid. Outliers in the training set that do not affect the mean point of training set are purged in the work of GW Gates [6], thus reduce the reorganization rate of sample in large datasets. One of the major pitfall of kNN i.e. memory limitation was addressed by Angiulli[7], by eliminating data sets which show similarity in training set.Guo and Bell in [8]proposed a model to automatic reorganize the k value in dynamic web pages of large web repositories. Automatic classification of data points in training set based on k value was a research finding of G Guo.In 2003 SC Baguai and K Pal assigned rank to training data for each category by using Gaussian distribution, thus it dominated the other variations of kNN[9]. Modified kNN[10] suggested by Parvin et.al in 2008 used weights and validity of data points to classify nearest neighbor and used various outlier elimination techniques. In the year, 2009 Zeng et.al planned a novel idea for kNN by using n-1 classes to the entire training to address the computational complexity [11]. The annoyed paper for this publication was of Z Youg work i.e. clusters formation using k nearest neighbors in text mining domains [12], proved as a robust method and overcome the defects of uneven distribution of training set.

Hubert et.al in 2013 composed an algorithm for correspondence search in the metric space of measure allocation in the thesaurus [13].In 2013, Du et.al proposed an effective strategy to accelerate the standard kNN, based on a simple principle: usually, nearer points in space are also, near when they areprojected into a direction, which is used to remove irrelevant points and decreases the computation cost[14].Basu et.al in 2013 assigned a document to a predefined class for a large value of k when the margin of majority voting is one or when a tie occurs. The majority voting method discriminates the criterion to prune the actual search space of the test document. This rule has enhanced the confidence of the voting process and it makes no prior assumption about the number of nearest neighbors[15]. One of the major drawbacks of k-Means clustering technique is its insightfulness to high-fliers; because it's mean value can be simply prejudiced by exterior values. The other partitioned clustering technique k-Medoids clustering which is a disparity of k-Means is more robust to noises and able-bodied to high-fliers. K-Medoids clustering is symbolized with an actual point as the center of the cluster instead of a mean point. The actual point is assumed as a cluster center, if it is located centrally and with minimum sum of distances to other points [16].

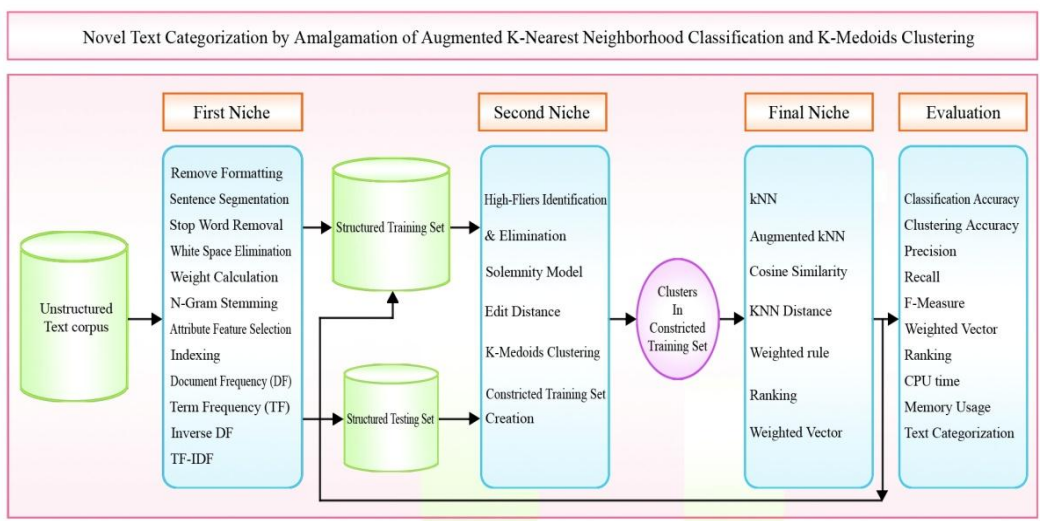

Figure 1. Roadmap of the amalgamative technique for text categorization

## 3. Proposed Work

The intuition of proposed algorithm operates in three niches. The first niche initiates the preprocessing on text mining corpus settled in the form of training set. This is accomplished by attribute feature selection to alienate the higher dimensions of training set and derive a constricted training set with minimal dimensions. The second niche uses the partitioned clustering k-Medoidsalgorithm to define k objects as a most centrally located points in the classes of constricted training set. A Solemnity model is designed to identify the cluster centers using edit distance as a similarity measure for each category in the training set. The worthiness of constricted training set is ensured by high-fliers identification and elimination. The final niche is embedded exclusively with an accurate text classification system using a novel augmentedkNN classifier with cosine text mining similarity measure, thus enhances the confidence of assigning a test sample to a trained categorical class. This insight the text classifier criterion to assign considerable weights to each object in the new constricted training set base upon their minimal average dissimilarity with the latest computed centroids. It was proven from the results attained that, the amalgamation of all the three niches rationale a better solution for evaluating the effectiveness of the text classification. The overview of the proposed intuition is outlined as figure 1 roadmap of the amalgamative technique for text categorization.

### 3.1. First Niche – Preprocessing of Text Corpus:

The preprocessing of text classification over datasets includes many key technologies before extracting the features from the text.The data cleaning process involves removing formatting, converting the data into plain text, segmentation on sentence, stemming [17], removing whitespace, uppercase characters, stop words removal [18] and weight calculation.The text classification is made simpler by eradication of inappropriate dimensions in the dataset. This was defensible by dispensation of the curse of dimensionality [19] i.e. higher dimensional data sets are reduced into lower dimensional data set without significant loss of information. In general,attribute feature selection methods are used to rank each latent feature according to a fastidious feature selection metric, and then engage the best k features. Ranking entails counting the incidence of each feature in training set as either positive or negative samples discretely, and then computing a function to each these classes.

### 3.1.1. Feature selection

Feature selection methods only selects as subset of meaningful or useful dimensions specific to applications from the original set of dimensions. The text mining domain, attribute feature selection methods include Document Frequency (DF), Term Frequency (TF), Inverse Document Frequency (IDF) etc. These methods fort requisites based on numerical procedures computed for each document in the text corpus.

The documents in the proposed work use the vector space model to represent document objects. Each document is symbolized by a vector $d$ in the term space such that $d = \{tf_1, tf_2, \ldots, tf_n\}$ where $tf_1, i = 1, \ldots n$ is the TF of the term $t_i$ in the document. To represent every document with the same set of terms, extractingall the terms found in the documents and use them, as the generated feature vector is necessary. The other method used was the combination of term frequency with inverse document frequency (TF-IDF). The document frequency $df_i$is the number of documents in anassortment of $N$ documents in which the term $t_i$transpires. Usually IDF is given as $\log(N/df_i)$. The weight of a term $t_i$ in a document is given by $w_i = tf_i * \log(N/df_i)$ [20]. To maintain theseattribute feature vector dimensionssensible, only $n$ terms with the maximum weights in all the documents are selected as n features. TF-IDF is termed as$TFIDF(t, d, D) = TF(t, d) * IDF(t, D)$, where $t$ is the terms in document $d$ and $D$ is the total number of documents in D.

Stemming [21]refers to the process of erasing word suffixes to retrieve the root (or stem) of the words, which reduces the complexity of the data without significant loss of information in a bag ofwords representing the text data. An attractive alternative to stemming used in this proposed work is character n-gramtokenization. Adamson and Borehamin 1974 introduced n-gram stemmersmethod[22] a conflating terms also called shared digram method. A digram is a pair of consecutive letters. In this approach, n-gram method is used as association measures to calculate distance between pairs of terms based on shared unique digrams. Formerly the distinctive digrams for the word pair are articulated and counted, with a similarity quantifier called Dice's coefficient for computation. Dice's coefficient is termed as $S = 2C/A + B$, where $A$ is the number of distinctivedigrams in the first word, $B$ the number of distinctivedigrams in the second, and $C$ the International Journal of Computational Science and Information Technology (IJCSITY) Vol.1, No.4, November 2013number of distinctivedigrams shared by $A$ and $B$. This similarity quantifieris used to bentall pairs of terms in the database, and also used to figure a similarity matrix.Since Dice coefficient is symmetric, a lower triangle similarity matrix is used.At the end of first niche the text preprocessing is well-rounded by building a strong bottomline and preparedness for accurate text classification.

## 3.2. Second Niche – High-filers Identification, Elimination & k-MedoidsClustering

High-fliers are the noisy and redundant data present in a data set, they materialize to be conflicting with the other residues in the data set. Such high-fliers are to be eliminated because they are the applicant for abnormal data that unfavorably guide to sculpt vagueness, unfair constraint inference and inaccurate domino effect. In text classification, High-fliers raise between each category because of the uneven distribution of data, as which carries to that the distance between samples in the same category to be larger than distance between samples in different categories. This is addressed by building a Solemnity model in k-Medoids clustering algorithm.

### 3.2.1. Solemnity Model

To reduce the complexity of constricted training set, and overcome the uneven distribution of text samples the k-Medoids clustering algorithm is used. The similarity measure used in identifying the categories in training set was Levenshtein distance (LD) or Edit distance [23]. This quantification is used a string metric for assessing the dissimilarityamidby the two successions. Edit distance between two strings $a, b$ is given by $lev_{a,b}(|a|, |b|)$ where

$$lev_{a,b}(i,j) = \begin{cases} \max(i,j) \; if \; min(i,j) = 0, \\ \min \begin{cases} lev_{a,b}(i-1,j) + 1 \\ lev_{a,b}(i,j-1) + 1 \\ lev_{a,b}(i-1,j-1) + [a_i \neq b_j] \end{cases} \end{cases} \text{otherwise,}$$

LD has numerous upper and lower bounds. If the comparing strings are identical, it results zero. If the strings are of the identical size, the Hamming distance is an upper bound on the LD distance. The LD between twostrings is no greater than the sum of their LD from a third string. At the end of the second niche, the high-fliers are eliminated and a solemnity model is built using k-Mddwith LD similarity measure and the training set is rebuilt by partitioning the objects into different categories, thus reduces the computational complexityand impacts on the memory usage tenancy. The cluster centers are treated as the k representative's objects in the constricted training set.

## 3.3. Final Niche – Appliance of AkNNand amending weights towards computed centroid

The second niche classifies the objects in the training phase by building aconstricted training set using k centroidsbut when an object from a testing set have to be classified, it is necessary to calculate similarities between object in testing set and existing samples in it constricted training sets. It then chooses k nearest neighbor samples, which have larger similarities, and amends weights using weighting vector.

To accomplish this model Cosine similarity is used as a distance measure in kNNtext classifier to classify the test sample towards the similarity with k samples in the training set. Cosine similarity distance measure is commonly used in high dimensional positive spaces, used to compare documents in text mining. IR uses this cosine similarity for comparing documents and to range values from 0 to 1, since TF-IDF weights cannot be negative.Each term is speculatively dispensed over a distinctive dimension and a vector is described for each document, where value of each

85

dimension matches to the number of times the term appeared in the document. The cosines of two vectors are resultant by using the Euclidean dot product formula $a.b = \|a\|\|b\|\cos\theta$. Assumed two vectors of attributes, A and B the cosine similarity, is symbolized using a dot product and magnitude as $similarity = \cos\theta = \frac{A.B}{\|A\|\|B\|} = \frac{\sum_{i=1}^{n} A_i * B_i}{\sqrt{\sum_{i=1}^{n}(A_i)^2} * \sqrt{\sum_{i=1}^{n}(B_i)^2}}$. The ensuing similarity vary from -1 (exactly opposite) to 1 (exactly the same), and with 0 (indicating independence). For text similarities, the attribute vectors A and B are usually the TF vectors of the documents

### 3.3.1. Assigning weights to samples

To assign precedence to the samples in constricted training set that have high similarity with the test sample the kNN classifier uses a distance weighted kNN rule proposed by Dudani[24]. Let $w_i$ be the weight of the $i^{th}$ nearest samples. The WkNN rule resolves the weight $w_i$ by using a distance function between the test samples and the $i^{th}$ nearest neighbor, i.e. samples with minor distances are weighted more profoundly than with the larger distances. The simple function that scales the weights linearly is $w_i = \begin{cases} 1 \text{ if } d_k = d_i \\ \frac{d_k - d_i}{d_k - d_1} \text{ if } d_k \neq d_1 \end{cases}$ where $d_i$ is the distance to the test sample of the $i^{th}$ nearest neighbor and the farthest $(k^{th})$ neighbor respectively. The WkNN rule uses the rank weighting function $w_i = k - i + 1$ to assign the weights. The proposed work used this function for computing the weights of kNNsamples belonging to individual classes.

The upshot observed at the end of the final niche was the AkNN classifies the objects in testing set with the distance weighted kNN rule compute ranks to each category and ranks k samples that have high similarities and accessory towards the centroids. The Inactive amalgamation of classification and clustering text categorization algorithm is obtainable as algorithm 1.

Algorithm1: Amalgamation of Classification and Clustering for Text Categorization

**Input:** A Raw training set composing of $n$ text documents $D = \{d_{i1}, d_{i2}, \ldots, d_{ii}, \ldots d_{in}\}$, a set of predefined $C$ category lables $C = \{C_1, \ldots, C_n\}$ and c clusters $c_1, \ldots, c_n$, and testing set sample documents $d$.

**First niche**

1: Preprocess the training set documents, $D = \{d_{i1}, d_{i2}, \ldots, d_{ii}, \ldots d_{in}\}$, into vectors from dataset
2: **For** each feature $f_1$ in the dataset **do**
3: Attribute feature selection by removing formatting, sentence segmentation, stop-word removal, white space removal, weight calcuation, n-gram stemming
4: Compute DF, TF, IDF and TF-IDF
5: **End for**

**Second niche**

6: Recognize and purge high-fliers by building a solmenity model using Edit distance
7: Obtain constricted training set with the preprocessed text document collection
8: **For** each processed document $d_i$ in D **do**
9: Compute c clusters $c_1, \ldots, c_n$, using K-Medoids clustering algorithm for all categories of $C$ $C_1, \ldots, C_n$ in D $\{d_{i1}, d_{i2}, \ldots, d_{ii}, \ldots d_{in}\}$ documents
10: For every cluster ID, generate and index the feature cluster vectors

11: Redistribute the constricted training set with cluster represetatives of each text categories.
**12: End for**

**Final niche**

13: **For** each processed document $d_i$ in D of constricted training set **do**
14: Classifythe testing set sample document $d$ using AkNN text classifer with Cosine Simalirity over the constricted training set $D$.
15: Assign the weights $w$ using distance weighted kNN rule to training set documents $D$ and assign rank to k samples that has the largest similarity and belongingness to $C$ categories.
16: Position the testing set document $d$ to the categories C which has the largest similairy and consign them with heavily weighted rank.
17: Build classifier on these mapped text document and record the classification accuray
**18: End for**

**Output:** Classifying testing documents $d$ in to $C$ categories with appropriate weight value $w$.

## 4. Experimental Analysis and Observations
### 4.1. Dataset

The referred algorithm uses Reuters-21578 a classical benchmark dataset for text categorization classification [25]. Reuters 21578 consist documents from the Reuters newswire, which was categorized and indexed in to grouping, by Reuters Ltd. and Carnegie Group in 1987. In 1991 David D. Lewis and Peter Shoemaker of Center for Informationand Language Studies, University of Chicago prepared and formatted the metafile for the Reuters-21578 dataset [26]. The experimental environment used is Intel core i3 CPU 530@2.93GHz 2.93 GHz, RAM 4GB, 32-bit Windows OS, and Java software. The software Package html2text-1.3.2 is used to convert the text corpus into text documents. The categories in text documents are separated as individual text documents using Amberfish-1.6.4 software.Thetext corpus consists of 21,578 documents, assigned with 123 different categories. Table 1 presents the initial distribution of all categories in Reuters-21578 text corpos before it is preprocessed.

Table1. Initial distribution of all categories in Reuters 21578

| Categories | #Documents | Categories | #Documents | Categories | #Documents |
|---|---|---|---|---|---|
| Earn | 3987 | Feed | 51 | Potato | 6 |
| Acq | 2448 | Rubber | 51 | Propane | 6 |
| Money | 991 | Zinc | 44 | Austdlr | 4 |
| Fx | 801 | Palm | 43 | Belly | 4 |
| Crude | 634 | Chem | 41 | Cpu | 4 |
| Grain | 628 | Pet | 41 | Nzdlr | 4 |
| Trade | 552 | Silver | 37 | Plywood | 4 |
| Interest | 513 | Lead | 35 | Pork | 4 |
| Wheat | 306 | Rapeseed | 35 | Tapioca | 4 |
| Ship | 305 | Sorghum | 35 | Cake | 3 |
| Corn | 255 | Tin | 33 | Can | 3 |
| Oil | 238 | Metal | 32 | Copra | 3 |
| Dlr | 217 | Strategic | 32 | Dfl | 3 |
| Gas | 195 | Wpi | 32 | F | 3 |
| Oilseed | 192 | Orange | 29 | Lin | 3 |
| Supply | 190 | Fuel | 28 | Lit | 3 |

| Sugar | 184 | Hog | 27 | Nkr | 3 |
|---|---|---|---|---|---|
| Gnp | 163 | Retail | 27 | Palladium | 3 |
| Coffee | 145 | Heat | 25 | Palmkernel | 3 |
| Veg | 137 | Housing | 21 | Rand | 3 |
| Gold | 135 | Stg | 21 | Saudriyal | 3 |
| Nat | 130 | Income | 18 | Sfr | 3 |
| Soybean | 120 | Lei | 17 | Castor | 2 |
| Bop | 116 | Lumber | 17 | Cornglutenfeed | 2 |
| Livestock | 114 | Sunseed | 17 | Fishmeal | 2 |
| Cpi | 112 | Dmk | 15 | Linseed | 2 |
| Reserves | 84 | Tea | 15 | Rye | 2 |
| Meal | 82 | Oat | 14 | Wool | 2 |
| Copper | 78 | Coconut | 13 | Bean | 1 |
| Cocoa | 76 | Cattle | 12 | Bfr | 1 |
| Jobs | 76 | Groundnut | 12 | Castorseed | 1 |
| Carcass | 75 | Platinum | 12 | Citruspulp | 1 |
| Yen | 69 | Nickel | 11 | Cottonseed | 1 |
| Iron | 67 | Sun | 10 | Cruzado | 1 |
| Rice | 67 | L | 9 | Dkr | 1 |
| Steel | 67 | Rape | 9 | Hk | 1 |
| Cotton | 66 | Jet | 8 | Peseta | 1 |
| Ipi | 65 | Debt | 7 | Red | 1 |
| Alum | 63 | Instal | 7 | Ringgit | 1 |
| Barley | 54 | Inventories | 7 | Rupiah | 1 |
| Soy | 52 | Naphtha | 7 | Skr | 1 |

In the first nitche preprocessing is imposed on the text corpos, further test the effectiveness and efficiency of Category specific classification data set is divided according to the standard Modified Apte split. This split results 13575 documents with training set, 6231documents with testing set and removes 1771 documents from Reuters-21578 documents since they are not used either in training or testing set. The utmost number of categories allocated to a document is 16 and its average value is 1.24.It also eliminates 47 categories that do not exist in both training and testings set, thus a total number of distinct terms in the text corpos after preprocessing is 20298. Table 2 list the top 12 categories with number of documents assigned both in training and testing sets after completion of preprocessing.

Table 2. Top 12 categories in Reuters-21578 after preprocessing

| Category | Training Set | Testing Set | Category | Training Set | Testing Set |
|---|---|---|---|---|---|
| Earn | 2870 | 1094 | Trade | 369 | 120 |
| Acquisition | 1659 | 170 | Interest | 338 | 142 |
| Money | 528 | 189 | Ship | 186 | 100 |
| Grain | 431 | 151 | Wheat | 221 | 82 |
| Crude | 389 | 189 | Corn | 171 | 67 |
| Earn | 2887 | 1097 | Trade | 386 | 119 |

**4.2 Evaluation Measures**

Information Retrieval (IR) is apprehensived with the directorialretrieval of information from a outsized number of textdocuments.There are two basic criterions widely used in document categorization field are precision and recall. Precision is the fraction of returned documents that are correcttargets, while recall is the fraction of correct target documents returned.

$$Recall = \frac{number\ of\ correct\ positive\ predictions}{number\ of\ positive\ examples},$$

$$Precesion = \frac{number\ of\ correct\ positive\ predictions}{number\ of\ positive\ predictions}$$

Van Rijsbergen [28] introduced the F1 measure as an estimation to to evaluate the text classification structure.It combines both recall and precision and is given as

$$F1 = \frac{2 \times Recall \times Precision}{(Recall + Precision)},$$

where $F1$ score reaches its best value at 1 and worst score at 0. Accuracy is defined as the percent of documents correctly classified in the classes based on the documents target labels.

The preprocessed text corpus was supplied as an input to the second nitche. The kMdd clustering algorithm groups the documents in to $k$ differeent categories. Table 3 summarizes the results after the application of k-Mdd clustering algorithm over the text corpus. Multiple category lables that tends to high-fliers are discarded and for clustering accurary uplifment clusters having less than 5 documents are also removed. The average number of categories per documentis 1.334 and the average number of documentsper category is about 151 or 1.39% of thecorpus.Table 4 gives the documents in cluster numbers 01 and cluster 06 and the set of categories associated with these documents. Table 5 summarizes the total number of clusters generated holding minimum and maximum number of text documents as cluster size. The k-Mdd clustering accuracy based on k value is given in Table 6. An important observation from Table 6wasas the k value increases a proportionate raise in computational time and plunge in accuracy is turned out.

Table 3. Results abridged after k-Medoids clustering on text corpus

| Description | Reuters text corpus | Description | Reuters text corpus |
|---|---|---|---|
| Total Number of documents | 21578 | Max. cluster size | 3854 |
| Number of documents used | 19806 | Min. cluster size | 5 |
| Number of documents Un-used | 1772 | Average cluster size | 171 |
| Number of clusters | 123 | Max. cluster size | 3854 |

Table 4. Sample Cluster numbers with document identifiers and categories

| Cluster Number | Document ID's | Categories |
|---|---|---|
| 01 | 12385, 12874, 12459, 16518, 18719, 2059, 3150, 3565, 3864,4335, 5805, 5625, 6211, 6481, 6501, 6744, 7994, 8050, 8997, 8449, 8333, 9121, 9381, 9355, 9457 | ship, earn, gas, strategic, metal, acq,grain ship, dlr, trade, money,fx |
| 06 | 12664, 12378, 14712, 2539, 456, 2564, 3751, 5421, 6789, 64561, 6781, 6977, 7947, 8239 | earn, gas, strategic, metal, acq, tin |

At the end of second nitche the high-fliers are eliminated and training set text corpus is clustered into different categories based on k-value. The ingrained final nitche is sequenced by triggering

International Journal of Computational Science and Information Technology (IJCSITY) Vol.1, No.4, November 2013the augumented kNN over training set and regaling solictations to testing set. The weighed vector using distance weighted kNN rule is imposed over the testing set based on high similarites and accessory towards the centroids for each categories in testing set.

It ranks the k nearest neighbors from the training set. The performance of AkNN text classifier with and with out weighting vector using evaluation criterion and comparisions with traditonal kNN is summarized as Table 7.

Table 5. Number of Clusters shaped with sizes

| Size of clusters (No.of documents) | Total No. of Clusters | Size of clusters (No.of documents) | Total No. of Clusters | Size of clusters (No.of documents) | Total No. of Clusters |
|---|---|---|---|---|---|
| 0-5 | - | 61-65 | 4 | 176-200 | 3 |
| 6-34 | 4 | 66-75 | 7 | 201-300 | 7 |
| 35-40 | 5 | 76-85 | 5 | 301-400 | 1 |
| 41-45 | 8 | 86-95 | 2 | 401-500 | 1 |
| 46-50 | 7 | 96-125 | 7 | 501-600 | 2 |
| 51-55 | 9 | 126-150 | 3 | 601-700 | 2 |
| 56-60 | 6 | 151-175 | 1 | **Total Clusters** | **84** |

Table 6. K-medoids clustering accuracy on text corpus based on k value.

| k-Value | Accuracy(%) | CPU time (sec) | k-Value | Accuracy(%) | CPU time (sec) |
|---|---|---|---|---|---|
| 1 | 0.9145 | 2.58 | 7 | 0.6230 | 08.43 |
| 2 | 0.9622 | 2.63 | 8 | 0.5724 | 09.73 |
| 3 | 0.7750 | 3.13 | 9 | 0.6185 | 24.28 |
| 4 | 0.7586 | 3.86 | 10 | 0.6573 | 30.65 |
| 5 | 0.6711 | 4.98 | 15 | 0.6091 | 56.53 |
| 6 | 0.7162 | 6.51 | 20 | 0.5458 | 57.26 |

Table 7. Preformance evaluation of AkNN text classifier on top 10 categories of Reuters-21578

| Categories | Precesion | | | Recall | | | F-Measure | | |
|---|---|---|---|---|---|---|---|---|---|
| | kNN | AkNN without Weights | AkNN with Weights | kNN | AkNN without Weights | AkNN with Weights | kNN | AkNN without Weights | AkNN with Weights |
| Acq | 0.89 | 0.91 | 0.93 | 0.93 | 0.94 | 0.95 | 0.9096 | 0.9248 | **0.9399** |
| Crude | 0.70 | 0.74 | 0.77 | 0.88 | 0.88 | 0.89 | 0.7797 | 0.804 | **0.8257** |
| Earn | 0.91 | 0.91 | 0.93 | 0.94 | 0.95 | 0.95 | 0.9248 | 0.9296 | **0.9399** |
| Grain | 0.78 | 0.81 | 0.84 | 0.81 | 0.84 | 0.87 | 0.7947 | 0.8247 | **0.8547** |
| Interest | 0.71 | 0.74 | 0.79 | 0.80 | 0.81 | 0.83 | 0.7523 | 0.7734 | **0.8095** |
| Money | 0.60 | 0.64 | 0.69 | 0.87 | 0.87 | 0.91 | 0.7102 | 0.7375 | **0.7849** |
| Fx | 0.77 | 0.81 | 0.88 | 0.85 | 0.84 | 0.87 | 0.808 | 0.8247 | **0.875** |
| Ship | 0.66 | 0.74 | 0.79 | 0.89 | 0.89 | 0.93 | 0.7579 | 0.8081 | **0.8543** |
| Trade | 0.52 | 0.54 | 0.57 | 0.67 | 0.71 | 0.74 | 0.5855 | 0.6134 | **0.644** |
| Wheat | 0.63 | 0.66 | 0.71 | 0.31 | 0.35 | 0.37 | 0.4155 | 0.4574 | **0.4865** |

The momentous observation on F-measure from Table 7 proliferates AkNN with Weight as the best performer, despite frantic attempts made by the enterent methods. Figure 2 exhibits the F-Measure dominance of weighted AkNN of Text classifiers over top 10 categories in Reuters text corpus. The text classifiers overall accuracy and CPU computational efficiency of AkNN with and without weighting vector is compared with the traditional kNN and is shown as Table 8.

The significant surveillance from Table 8 was AkNN with Weight outperforms its competitor both in training and testing phases. The overall accuray of conventional classifier could not tug at the heartstring of white-collar expatriate of weighted vector AkNNs, thus it complains the pro-activeness of amalgmative technique for text mining.

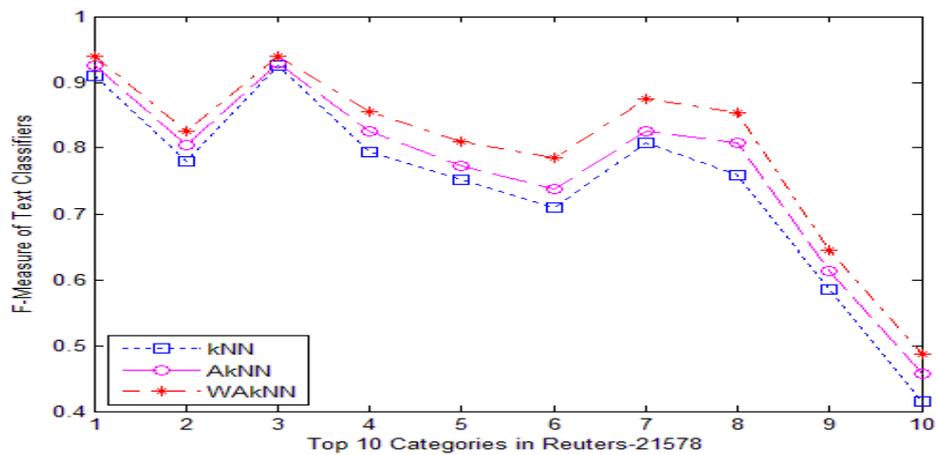

Figure 2. F-Measure of Text classifiers over top 10 categories in Reuters-21578 text corpus

Table 8. Comparision of Algorithms CPU Computational Time and accuracy

| Modified Apte split | AkNN with Weight | AkNN without Weight | kNN |
|---|---|---|---|
| Training Set | **21.6 min** | 58.47 min | 1.50 hrs |
| Testing Set | **15.4 secs** | 51.42 sec | 1.54 sec |
| **Text Classifiers overall Accuracy(%)** | **0.942** | 0.951 | 0.873 |

## 5. Conclusion

Due to the exponential increase in huge volumes of text documents, there is a need for analyzingtextcollections, so several techniques have been developed for mining knowledge from textdocuments. To strengthen the expressiveness of text classification, a novel amalgamative method using AkNN text classifier and kMdd clustering is tailored. It was well suited for dimension reduction, attribute feature selection, high-flier identification and reduction, and assigning weights/ ranks to k nearest neighbors using weighing vector. The experimental analysis conducted on the benchmark dataset Reuters-21578 rationalizes the reference method substantially and significantly surpasses traditional text categorization techniques, thus makes the amalgamation method a very promising and easy-to-use method fortext categorization.

Further research may address the issue of reduction of computational cost, by using soft computing approaches for text classification and clustering method. There is a scope to investigate the adjustments in the weighting vector based on k- value in accordance to the size of the categories and their belongingness to text classifier, thus speculating fruitful success in deriving accuracy and high performance over training and testing set.

**Authors**

**RamachandraRaoKurada** is currently working as Asst. Prof. in Department of Computer Applications at Shri Vishnu Engineering College for Women, Bhimavaram. He has 12 years of teaching experience and is a part-time Research Scholar in Dept. of CSE, ANU, Guntur under the guidance of Dr. K KarteekaPavan. He is a lifemember of CSI and ISTE. His research interests are Computational Intelligence, Data warehouse & Mining, Networking and Securities.

**Dr. K. KarteekaPavan** has received her PhD in Computer Science &Engg. fromAcharyaNagarjuna University in 2011. Earlier she had received her postgraduate degree in Computer Applications from Andhra University in 1996. She is having 16 years of teaching experience and currently working as Professor in Department of Information Technology of RVR &JC College of Engineering, Guntur. She has published more than 20 research publications in various International Journals and Conferences. Her research interest includes Soft Computing, Bioinformatics, Data mining, and Pattern Recognition.